\documentstyle[12pt,epsf,fleqn]{elsart}

\def \ni{\noindent}

\def \be {\begin{equation}}
\def \ee {\end{equation}}

\begin{document}

\ni {\bf Corresponding author}\\
Prof. Dr. H.V. Klapdor-Kleingrothaus\\
Max-Planck-Institut f\"ur Kernphysik\\
Saupfercheckweg 1\\
D-69117 HEIDELBERG\\
GERMANY\\
Phone Office: +49-(0)6221-516-262\\
Fax: +49-(0)6221-516-540\\
email: klapdor@gustav.mpi-hd.mpg.de\\

\begin{frontmatter}
\title{A digital multi-channel spectroscopy system with 100 MHz flash ADC 
module for the GENIUS-TF and GENIUS projects.}

\author{T. Kihm, V.F. Bobrakov, H.V. Klapdor-Kleingrothaus}
\address{Max-Planck-Institut f\"ur Kernphysik, PO 10 39 80, 
  D-69029 Heidelberg, Germany}

\date{}

\begin{abstract}
	In this paper we will present the first results 
	of applying a digital processing technology in low-level 
	gamma spectroscopy with HPGE detectors. An experimental 
	gamma spectrometer using Flash ADC module is built and tested. 
	The test system is now under development and shows major 
	advantages over the traditional analog technologies. 
	It will be installed for the GENIUS-TF and GENIUS projects 
	in Gran-Sasso in early 2003.
\end{abstract}
\end{frontmatter}

\section{Introduction}

	The GENIUS-TF project 
\cite{1} 
	and GENIUS project 
\cite{2} 
	- the latter being under construction in the Gran Sasso 
	Underground Laboratory - 
	require a new modular, multi-channel data acquisition system 
	and electronics, capable of taking data from up to 300 
	and more detectors simultaneously. Since last year 
	a new range of Flash ADC modules with high sampling rates 
	of 100\,MHz and resolution of 11-12 bits are available on the market. 
	With these devices it is now possible to develop a multi-channel 
	system for gamma-spectroscopy if the count range is in moderate 
	range. Using this approach it is possible to capture 
	the detailed shape of the preamplifier signal with high speed ADC, 
	and then perform digitally all essential data processing functions, 
	including precise energy measurement over a range of 3\,keV-3\,MeV, 
	rise time analysis, ballistic deficit correction and pulse shape 
	analysis. Thus we can obtain both the energy and the pulse shape 
	information from one detector using one channel 
	of the Flash ADC module.



\section{Test detector and 8-channel electronics}

\begin{figure}[b]
\epsfysize=70mm\centerline{\epsffile{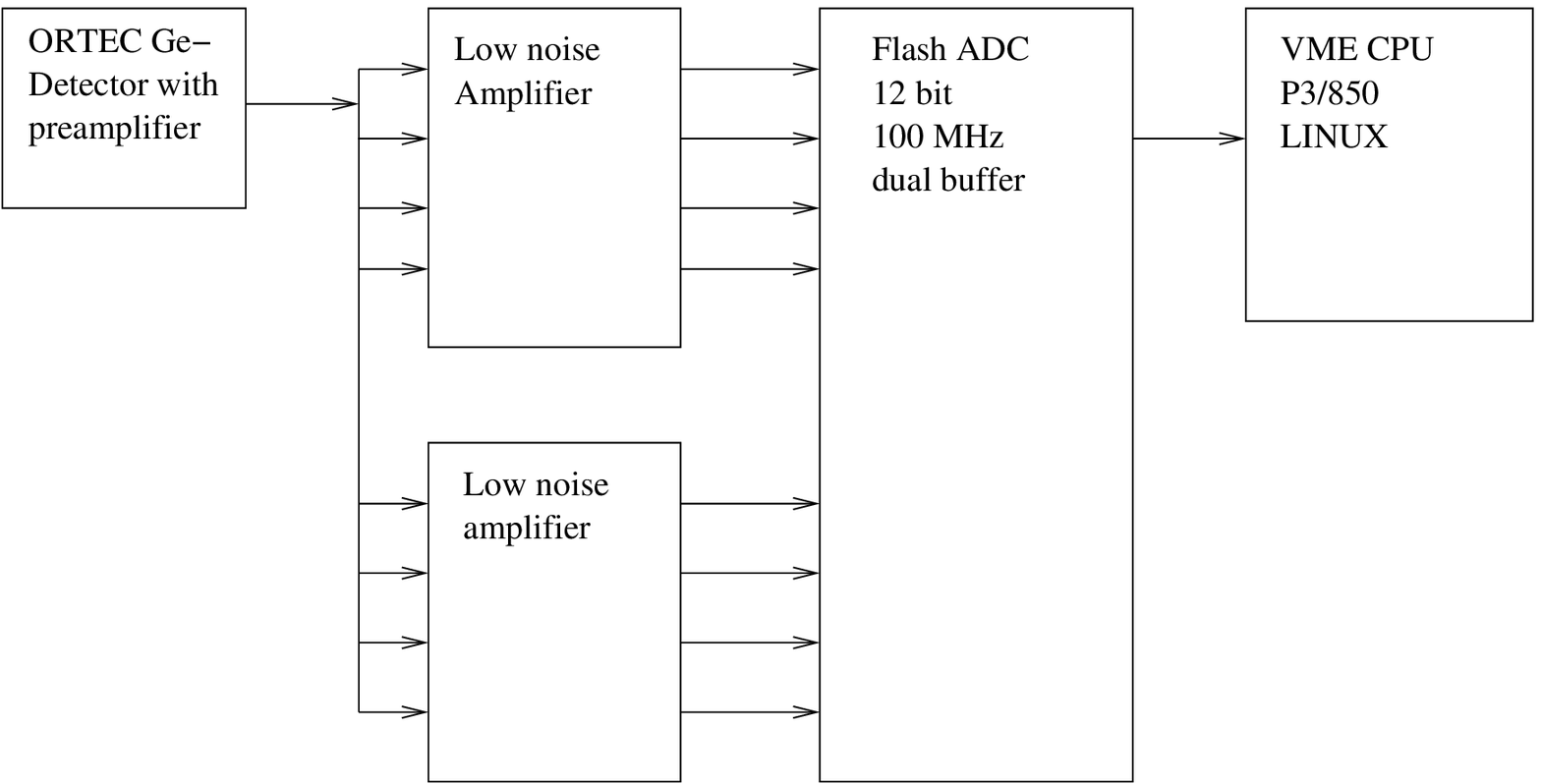}}
\caption{Schematic diagram of the 8-channel test system. 
	An output of the single Ge detector is split and connected 
	to eight inputs of the spectrometer for test purpose.}
\label{Fig1}
\end{figure}

	To test the digital processing approach for gamma spectroscopy 
	the following setup has been used:

	An ORTEC germanium detector with a built-in low noise 
	preamplifier was connected to two 4-channel low noise amplifiers 
	developed by MPI f\"ur Kernphysik in Heidelberg. 
	The amplifier/filter has a variable gain ranging from 10 to 100, 
	a variable offset with a maximum of 4\,V, a variable differential 
	filter, a pole zero cancellation circuit and two integral filters. 
	Currently we do not use any analog shaping at this stage, 
	these amplifiers provide only the necessary gain and DC offset. 
	The signal outputs from these amplifiers were directly connected 
	to the Flash ADC module developed by SIS (model 3300, 100\,MHz 
	sampling frequency, 12 bit, 8 channels, dual buffer). 
	A VME CPU with Intel P3/850\,MHz CPU, running Linux 
	as operating system, reads out the Flash ADC. 
Fig.\ref{Fig1} 
	shows a schematic diagram of the used test system.

	In the current test system the VME CPU waits for 
	at least 10 events (2000 samples for every event) sampled 
	in the dual buffer memory. The trigger for each event 
	is produced by the flash ADC, which has a trapezoid 
	FIR filter and a digital threshold circuit 
\cite{4} 
	for each of the 8 channels. 
	After the 10 events have been triggered the memory banks 
	of the FADC are switched and the CPU reads out the non-active 
	memory bank. Due to this double buffer feature of the FADC 
	we can achieve nearly zero dead time 
	(less than 10\,$\mu$sec for 10 events) as long as the buffers 
	of the FADC 
	are not overrun. After the read out which is currently available 
	with 4\,Mbytes/sec the test system applies digital filters 
	to the sampled data to compute the energy, the rise time, 
	and the timing of the input pulse.



\section{Digital signal processing for gamma spectroscopy}

\begin{figure}[b]
\epsfysize=90mm\centerline{\epsffile{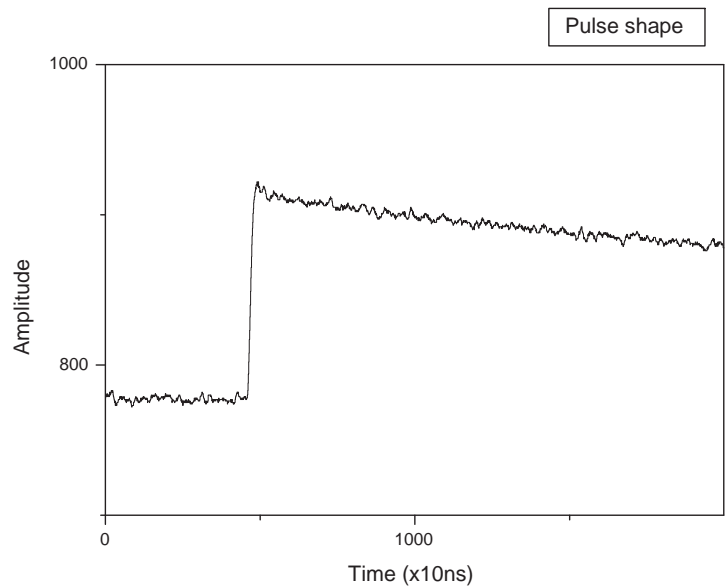}}
\caption{Typical Ge detector pulse recorded by Flash ADC.}
\label{Fig2}
\end{figure}

\begin{figure}[!htp]
\epsfysize=90mm\centerline{\epsffile{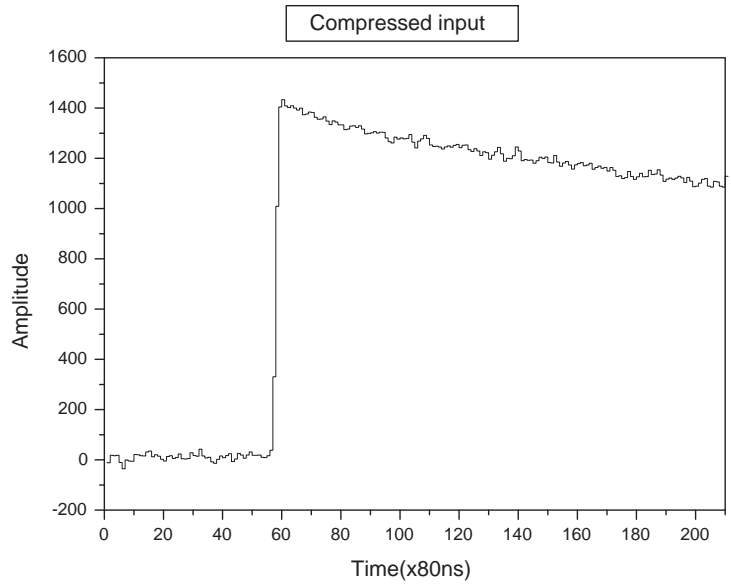}}
\caption{Compressed input pulse. A base line is subtracted.}
\label{Fig3}
\end{figure}

	Each analog stage in a conventional spectrometer introduces 
	irreversible signal distortion that cannot be removed 
	by further processing stages. If the input signal is digitized 
	at early stage, we can eliminate this problem and obtain 
	the best result using a digital processing.
	In our test system each pulse from the Ge detector is digitized 
	at 100\,MHz and stored as 2000 16-bit words in the CPU memory (see 
Fig. \ref{Fig2}). 
	The first 400 words are used for base line calculation. 
	After subtracting the base line, we compress the 2000 samples 
	to 250 by summing every 8 samples (see 
Fig. \ref{Fig3}), 
	before applying a digital filter.

	As a first stage of digital filtering we use a differential 
	filter with P/Z compensation :
 
\begin{equation} \label{formula_rate}
y[n]=a_0x[n] + a_1x[n-1] + b_1y[n-1] + px[n-1].
\end{equation}        

	A simple recursive filter is used as one stage of n-pole 
	integral filter:
 
 \begin{equation} \label{formula_rate}
y[n]=a_0x[n] + b_1y[n-1].
\end{equation}        
                 			 
	Repeating these calculations n times we obtain the n-pole 
	integral filter. The coefficients are found from these 
	simple equations 
\cite{3}.

	Differential filter:
	\begin{center}
	$a_0 = (1 +x)/2$
       
	$a_1 = - (1 +x)/2$       
 	
	$b_1 = x$	
	\end{center}

	Integral filter:
	
\begin{center}
	$a_0 = (1 - x)$
       
	$b_1 = x$	
	\end{center}
		
	The value for $x$ can be found from the desired time constant $d$ 
	of the filter:

\begin{center}
	$x = exp{({-1}/{d})}$
	\end{center}

	The coefficient $p$ in formula (1) is intended for pole/zero 
	compensation of the preamplifier RC time constant. 
	The optimal value for $p$ can be found experimentally 
	for every particular preamplifier/detector combination.

\begin{figure}[!htp]
\epsfysize=90mm\centerline{\epsffile{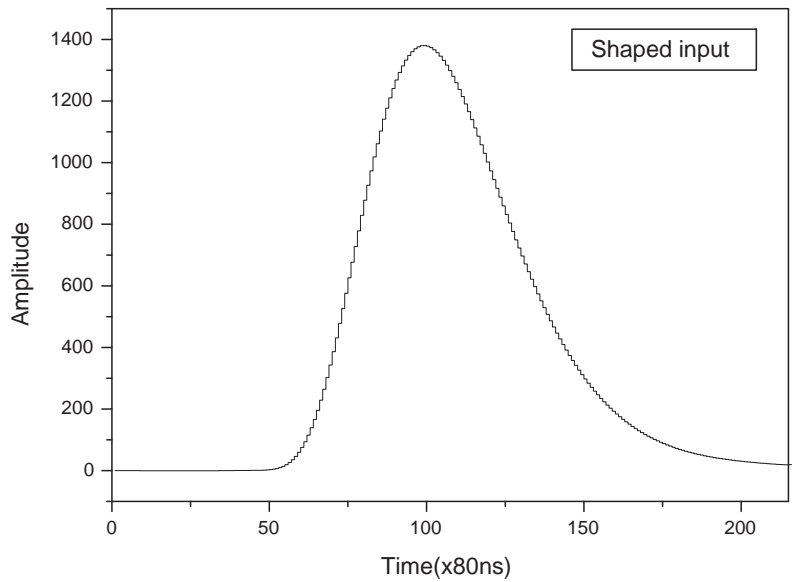}}
\caption{The shape of input pulse after applying a differential 
	filter with P/Z compensation and 8-pole integral filter.}
\label{Fig4} 
\end{figure}

	The energy value for a given event is calculated as a peak 
	value of the filtered pulse sample (see 
Fig.\ref{Fig4}). 
	The energy value is used 
	for an energy spectrum histogram filling.



\section{Experimental results.}

	The system was extensively tested for a 3 months period. 
	After optimization of the parameters we have found that 
	the best energy resolution was obtained for the 8-pole digital 
	filter with a time constant of 1.6\,$\mu$sec. With this filter 
	the measured resolution of the ORTEC germanium detector 
	was 1.9\,keV (FWHM) for the 1332\,keV  $^{60}$Co line, and 1.3\,keV 
	for 59.5\,keV  $^{241}$Am line.

\begin{figure}[!htp]
\epsfysize=90mm\centerline{\epsffile{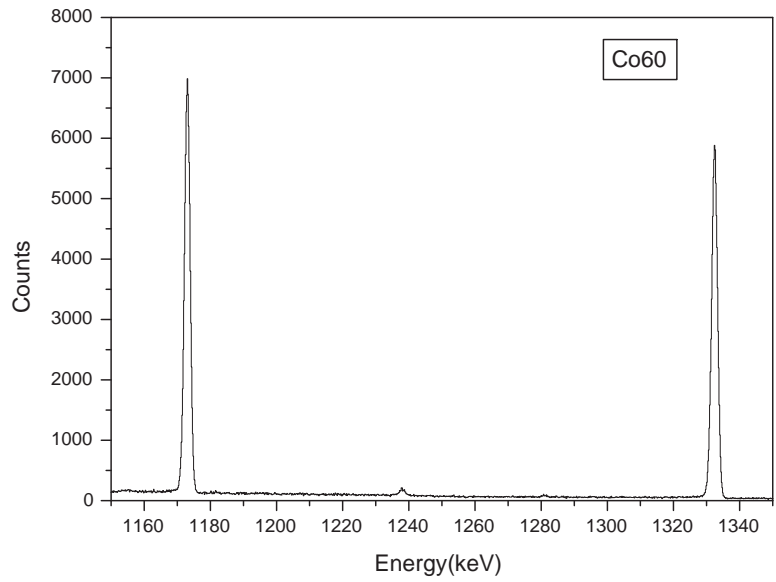}}
\caption{Spectrum measured with  $^{60}$Co source.}
\label{Fig5}
\end{figure}

	The resulting energy spectrum of the  $^{60}$Co 
	test source is presented in 
Fig. \ref{Fig5}.

	The energy resolution has nearly reproduced the FWHM measured 
	with a peak sensing ADC (1.8\,keV).

\begin{figure}[!htp]
\epsfysize=90mm\centerline{\epsffile{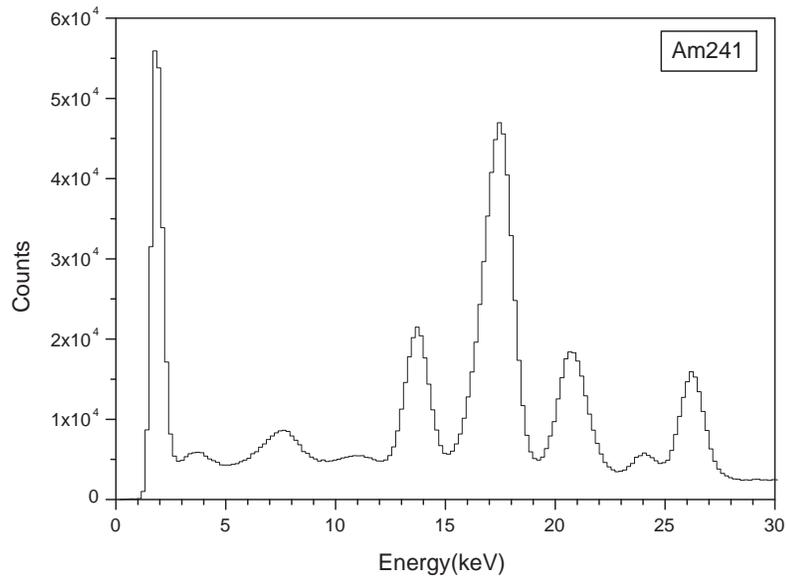}}
\caption{Low-energy part of the spectrum measured with an  $^{241}$Am source.}
\label{Fig6}
\end{figure}

Fig. \ref{Fig6} shows the low-energy part of the  $^{241}$Am spectrum. 
	We have found that the low-energy threshold of the test system 
	is about 3\,keV for the given detector. It will be lower  
	for a detector with a better energy resolution. 
	In any case it is already a satisfactory value 
	for the GENIUS-TF facility 
\cite{1}. 
	The measured integral non-linearity of the test system 
	is less than 0.01\%, which is corresponding to the integral 
	non-linearity of the used ADC. The measured differential 
	non-linearity is less than 1\%. 
	To determine the long time stability, we tested our system 
	for a three weeks period with a $^{60}$Co source. 
	The air temperature in the laboratory was 21$\pm$1$^0$C. 
	At this condition the maximum peak shift was less 
	than 0.5 channels ( 0.007\% ). 
	The maximum count rate of the test system is 4000 events per second.


\section{Conclusion}
	A digital spectroscopy system has been developed for 
	the GENIUS-TF and GENIUS projects. Using a digital 
	signal processing instead of analog reveals better accuracy 
	and stability limited only by ADC quality. Our tests have proved, 
	that this system shows an ultimate performance making 
	it the best choice for low-level gamma spectroscopy. 
	The system will be installed in the Gran-Sasso underground 
	lab in the beginning of the year 2003.



\section{Acknowledgement}

	The authors thank Mr. C. D\"orr for his help 
	in the beginning of this project.
	The authors would like to thank Prof. V.A. Bednyakov 
	for his permanent interest in this work.




\clearpage


\end{document}